# Rethinking the protein folding problem from a new perspective

Jorge A. Vila

IMASL-CONICET, Universidad Nacional de San Luis, Ejército de Los Andes 950, 5700 San Luis, Argentina.

**Abstract**

One of the main concerns of Anfinsen was to reveal the connection between the amino acid sequence and their biologically active conformation. This search gave rise to two crucial questions in structural biology, namely, *why* the proteins fold and *how* a sequence encodes its folding. As to the *why*, he proposes a plausible answer, namely, at a given milieu a protein folds into its functional form -native state- because such structure represents the lowest free-energy minimum among all feasible conformations -the thermodynamic hypothesis or Anfinsen's dogma. As to the *how*, this remains as an unsolved challenge and, hence, this inquiry is examined here from a new perspective of protein folding, namely, as an 'analytic whole' -a notion proposed by Leibnitz and Kant's. This new perspective forces us to discuss in detail why the theoretical force-field-based approaches have failed in both their ability to predict the three-dimensional structure of a protein accurately and in their capacity to answer one of the most critical questions in structural biology: *how* a sequence encodes its folding. It is worth noting that the problem of accurately determining the three-dimensional structure of a protein -for a given amino acid sequence- is considered to have been solved, *viz.*, by either state-of-the-art numerical methods or by experimental methods. Therefore, the pros and cons of each of these approaches nor a relative comparison of their precisions will not be discussed here.

There is overwhelming evidence showing that state-of-the-art numerical methods can predict, with high accuracy, a three-dimensional (3D) structure of a protein (Tunyasuvunakool *et al*., 2021; Kryshtafovych *et al*. 2021; Marx, 2022). In spite of this, *how* a protein amino-acid sequence 'encodes' its folding -Anfinsen's challenge- is yet unknown. Consequently, numerous questions remain to be answered, *e.g*., an accurate determination of structural and marginal-stability changes upon protein point-mutations and/or post-translational modifications (Pancotti *et al*., 2022; Serpell *et al*., 2021; Buel *et al*., 2022). Yet, all the above downsides do not belong to the state-of-the-art numerical methods alone. Indeed, neither X-ray crystallography, cryogenic electron-microscopy (Cryo-EM), nor NMR spectroscopy provide an answer to Anfinsen's challenge despite all methods needing to know the amino acid sequence in advance. This is certainly not surprising, because none of these numerical/experimental methods was designed to answer the question but to provide an accurate prediction/determination of the tridimensional structure of the protein. Therefore, the search for an accurate answer to the query of *how* the protein amino-acid sequence encodes its folding is a problem that transcends the information provided by any of the existent methods to predict/determine the protein 3D structure. This gives rise to one fundamental conjecture to solve Anfinsen's challenge: the protein folding problem should be conceived as an 'analytic whole'. This statement arises from Leibniz & Kant's notion of space (and time), devised as 'analytic wholes', *i.e*., the one in which "*…its priority makes it impossible to obtain it by the additive synthesis of previously existing entities…*" (Gómez, 1998). From this point of view, methods based mainly on additive pairwise interactions may not give a precise answer to Anfinsen's challenge because such methods consider the 'whole' as *a posteriori* rather than as *a priori*. Therefore, the solution demands solving an *n*-body problem, with *n* being the number of amino acids in the sequence. The latter seems to be a necessary condition to solve the protein folding problem. This demand for treating the protein folding as an 'analytic whole' is analogous to that needed by numerical/experimental methods aimed at predicting/determining the protein-tridimensional shape, namely, the existence of the structure as a 'whole' as an *a priori* for its resolution. All of the above gives rise to the following theoretical thought. For more than ~60 years, the protein folding problem has been unsuccessfully attempted to be solved at the atomic level, except for a few exceptions (Kussell *et al*., 2002; Vila *et al*., 2003; Lindorff-Larsen *et al*., 2011), by using (force) 'fields' that are defined, beyond details, by an additive sum of pairwise interactions (Arnautova *et al*., 2006; Best, 2019). Then, can we conclude that this has been a failure of the 'field' concept? -that is the most influential discovery since Newton's time, which was crucial for success in formulating physical-major problems like Maxwell's equations, or the theory of relativity (Einstein & Infeld, 1961). The answer is undoubtedly no, and the reason is the following. The failure does not arise from the 'field' concept, but from the assumption that the 'field', around and between atoms, can be accurately described solely by pairwise additive (two-body) interactions. The latter implies leaving out the *k*-body interactions ($2 < k \leq n$), despite all the existing pieces of evidence showing their relevance for an accurate solution to protein folding (Ejtehadi *et al*., 2004; Wang *et al*., 2021). That reductionism in the application of the 'field' concept should not surprise us. It is standard practice in science to use the simplest possible way to solve a problem, especially if the chosen strategy has previously led to successful predictions on similar problems. The successful Zimm-Bragg (1959) nearest-neighbor approximation to study the thermal-induced helix-coil transition -by the transfer-matrix method (Poland & Scheraga, 1970)- is a good example of the latter. Indeed, its transferability to analyze the pH-induced helix-coil transition in charged macromolecules (Zimm & Rice, 1960) leads to untreatable matrix analysis - from a theoretical point of view- because the long-range nature of the electrostatic interactions demands consideration of *all* charge-charge interactions, no matter the position of the ionizable groups in the sequence. If this condition is not fulfilled, a good result will always be suspected of being fortuitous (Liem *et al*., 1970). Therefore, options -to the transfer matrix- were a necessary condition to get an analytical and precise solution to the ionizable homopolypeptide (poly-L-lysine) problem, *e.g*., by using statistical mechanical analysis such as the one employed to study a one-dimensional lattice gas in which the pairwise potential is exponential and repulsive, namely, a Kac-type potential rather than a Debye-Hückel one (Vila, 1987). This alternative standpoint enabled us to crack the problem at the expense of leaving many details of the system aside, maintaining the long-range character of the electrostatic interactions, *i.e*., without trimming charge-charge interactions (Vila, 1986a, 1986b). This example teaches us a lesson. The existence of connections between similar systems might provide us with the necessary tools to solve problems that otherwise may remain unsolved. Yet, we should be very careful in the search for such associations. For example, a Coulombian and Newtonian potential has the same functional form, specifically $\propto 1/r$, with *r* being the distance between two bodies; hence, we could be tempted to think of the ionizable homopolypeptide problem as another sort of *n*-body gravitational-problem. However, as discussed above, an accurate solution for the ionizable homopolypeptide problem was obtained by a 'sum of parts' approach, *i.e*., a sum of pairwise additive interactions, rather than by treating it as an *n*-body problem. The reason for the latter lies in the fact that the homopolypeptide (poly-L-lysine) is a sequence of chemical-bonded identical amino acids -under a Coulombian repulsive

interaction- rather than a system containing different masses without any bond among them other than their mutual gravitational attraction, *i.e*., a collection of bodies obeying the celestial mechanic's law. This brief illustration gives rise to a new question. What would happen if the peptide sequence contained different amino acids? If this were the case, then the homopolypeptide would become a polypeptide and, hence, the above approach of analyzing its folding in terms of a 'sum of parts' might no longer be valid, *e.g*., for polypeptides able to form thermodynamically stable tridimensional structures -such as proteins. This example leads us to a second conjecture. If the local interactions -between nearest-neighbor residues- are or are not enough to accurately determine the thermodynamically most stable conformations for a polypeptide will define whether a (force) 'field' should be defined as a 'sum of parts' -a *two*-body problem- or as an 'analytic whole' -$n$-body problem. Indeed, there are numerous pieces of evidence that a fairly-accurate prediction of polypeptide conformations is possible by using all-atom force fields defined as a 'sum of parts' (Daura *et al*., 1999; Vila *et al*., 2000, 2001; Gnanakaran *et al*., 2003; Gnanakaran & García, 2005; Makowska *et al*., 2007; Georgoulia & Glykos, 2019; Dolenc *et al*., 2022); because the most relevant interactions are the local ones. On the contrary, if the non-local -tertiary- interactions are crucial to predict its thermodynamically most stable structure -such as for proteins- the use of such force fields doesn't generally lead to accurate-enough predictions. Proof of this has been the smashing success of state-of-the-art numerical methods (Marx, 2022). Despite this, Anfinsen's challenge remains unsolved. Hence, conceiving protein folding as an 'analytic whole' -a Leibniz & Kant's notion of space (and time)- implies that the (force) 'field' -between and around atoms- must be determined by solving an *n*-body problem. From this point of view, the 'field' would perform like the diffraction peak intensity -in X-ray crystallography- defined by the arrangement of atoms within the entire crystal (Ilari & Savino, 2008).

All in all, our analysis provides a strong argument as to why science needs philosophy (Laplane *et al*., 2019). In this regard, philosophy together with the history of science, constitute a complementary science "... *with the aim of improving scientific knowledge in ways that are not taken up by scientists themselves…*" (Chang, 2021). Indeed, the introduction of a classic philosophical notion (Gómez, 1998) enabled us to examine the capability of standard (force) fields -defined as a 'sum of parts'- to study polypeptide folding and, further, to analyze under which conditions this problem should be conceived from a new perspective; in other words, as an 'analytic whole', a Leibniz & Kant's notion of space (and time). Solving the protein folding problem following this approach is, no doubt, a daunting task. However, having had to wait for more than ~60 years to understand how a sequence codes their folding may well be a price to pay, and, perhaps, it may help to open new avenues for further research in the protein folding field.

**Acknowledgments**

The author acknowledges support from the IMASL-CONICET-UNSL and ANPCyT (PICT-02212), Argentina.